\documentclass[twocolumn,prl,showpacs,floatfix]{revtex4}

\usepackage{graphicx}

\begin{document}

\title
{
``Glassy" Relaxation in Catalytic Reaction Networks 
}

\author
{ 
Akinori Awazu
}

\affiliation
{
Department of Mathematical and Life Sciences, Hiroshima University,
Kagami-yama 1-3-1, Higashi-Hiroshima 739-8526, Japan.
}
    
\author
{ 
Kunihiko Kaneko
}

\affiliation
{
Department of Basic Science, University of Tokyo $\&$
ERATO Complex Systems Biology, JST,\\
Komaba, Meguro-ku, Tokyo 153-8902, Japan.
}

\date{\today}

\begin{abstract}
Relaxation dynamics in reversible catalytic reaction networks is studied, revealing two salient behaviors that are reminiscent of glassy behavior: slow 
relaxation with log(time) dependence of the correlation function, and emergence 
of a few plateaus in the relaxation. The former is explained by the 
eigenvalue distribution of a Jacobian matrix around the equilibrium state 
that follows the distribution of kinetic coefficients of reactions.  The 
latter is associated with kinetic constraints, rather than metastable states,  
and is due to the deficiency of catalysts for chemicals in excess and
negative correlation between the two chemical species.  
Examples are given, and generality is discussed.

\end{abstract}

\pacs{87.16.Yc, 82.39.Rt, 05.40.-a}

\maketitle

Cells are usually not in thermal equilibrium, and biological functions
are believed to operate under non-equilibrium conditions.  The relevance of 
non-equilibrium conditions to pattern formation has been discussed for 
decades\cite{Nicolis} since
the pioneering work of Schr\"{o}dinger\cite{Schro}. In contrast to physics and chemistry, however, 
such non-equilibrium conditions are not imposed
externally but have to be sustained by a biological system itself.
This sustainment might then suggest the existence of some bootstrapping
process in which biochemical reactions under non-equilibrium
conditions could suppress relaxation to equilibrium.  Even though this argument
may be too naive for currently known living organisms that adopt advanced mechanisms 
using cell membranes, it is nevertheless important in 
considering the origin of life.

In physics, the reluctance to relax to equilibrium has been studied in 
glass, and a certain complex free energy
landscape structure has been elucidated \cite{glass1,glass2,GR1,GR2}.  As
an alternative to such structural studies, kinetic mechanisms to
suppress the relaxation have recently been proposed \cite{Nakagawa-kk,
Morita, AK1, Shaw-Packard}.  'Kinetically constrained models'
have gathered much attention \cite{KCM1,KCM2,KCM3}, where the relaxation to 
equilibrium is slowed down due to a kinetic bottleneck.  
In the present Letter, we show that, in a system with a catalytic
reaction network, relaxation to thermal equilibrium is generally
slowed down due to a kinetic constraint.

We consider a network of reactions consisting of $M$
chemical components ($X_i$, $i=1,\cdots,M$), each of which is
catalyzed by one of the $M$ components.  Transformation between
chemicals $X_i$ and $X_j$ is catalyzed by $X_c(i,j)$, i.e.,
\begin{equation}
 X_i+X_c \rightleftharpoons^{k_{i,j}}_{k_{j,i}} X_j + X_c.
\end{equation}
The reaction network consists of the above reactions,
with the total number of reactions $G \geq M$. We assume that all
chemical species are percolated to any other through these
reactions.  The system is closed, without inflow of chemicals or energy 
from the outside.  Note that the number of molecules, accordingly 
$\sum_i x_i \equiv S$, is conserved by the above reactions,  where 
$x_i$ is the concentration of each chemical species $i$.

To assure the relaxation to thermal equilibrium,  the ratio
of forward to backward reactions is set so that it satisfies the
detailed balance condition.   It is satisfied by
allocating energy $E_i$ to each molecular species, and setting the ratio of
forward ($k_{i,j}$) to backward ($k_{j,i}$) reactions in eq. (1) to 
$k_{i,j} / k_{j,i} = \exp(-\beta (E_j-E_i))$, where $\beta$ is the 
inverse temperature.  As a result, the equilibrium concentration 
$x_i^{eq}$ satisfies $x_i^{eq}= s \exp(-\beta E_i)$ with 
$s = S (\sum_l \exp(-\beta E_l))^{-1}$.

Here we take a continuum description, so that the dynamics of the
concentration is given by the rate equation 
\begin{equation}
\dot{x_{i}} = \sum_{j,c} Con(i, j ; c) x_c(k_{j,i} x_{j} - k_{i,j}x_i), 
\end{equation} 
with \begin{math} k_{i, j} = \min\{1, \exp(-\beta (E_j-E_i))\} \end{math},
and $Con(i, j ; c) = Con(j, i ; c) = 1$ if there is a reaction path, as 
in eq. (1), and 0 otherwise\cite{note1}.  
Note that eq. (2) has a unique stable fixed point attractor $x_i^{eq}$,
without any metastable states.
We assume that the energy $E_i$ is distributed uniformly, as
$\frac{i}{M} \varepsilon$ ($\varepsilon$ is a constant)\cite{note2}. 
The network $Con(i, j ; c)$ is chosen
randomly by setting the average number of paths for each chemical
$K = 2G/M$.  As an example of a typical relaxation course, we set an
initial concentration with equal distribution over all chemicals,
i.e., the high-temperature limit (corresponding to $\beta=0$), and
study the evolution under given $\beta$.

\begin{figure}
\begin{center}
%\psbox[width=8.0cm]{}
%\includegraphics[width=8.0cm]{KA_FIG1.ps}
\includegraphics[width=8.0cm]{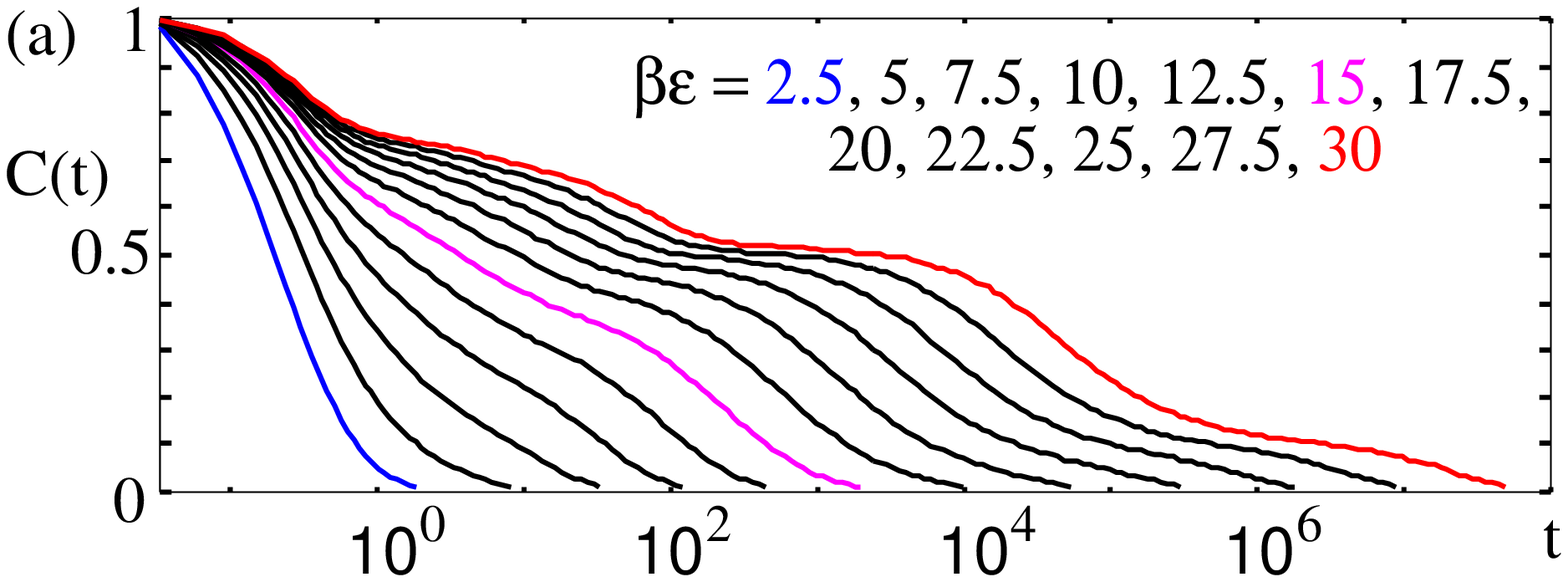}
\includegraphics[width=8.0cm]{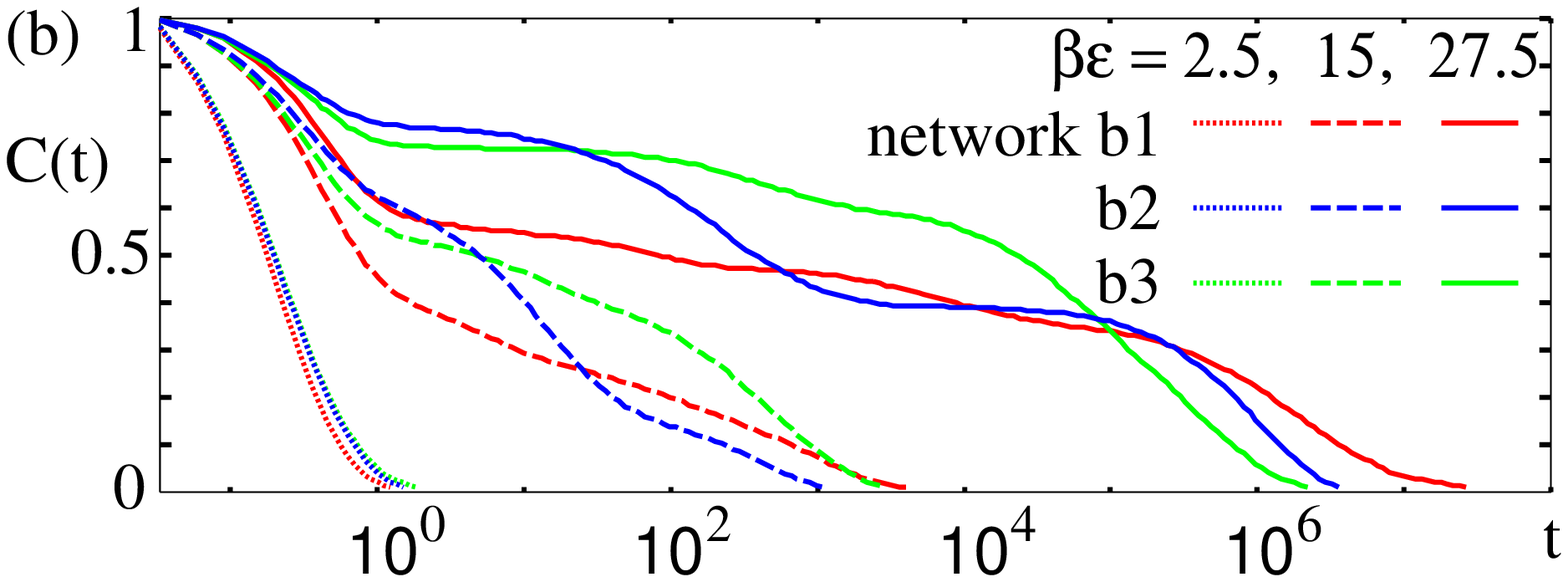}
\end{center}
\caption{(a)(b) Relaxation time course for four sets 
of networks ($M = 24$, $K=8$) for several $\beta$. }
\end{figure}

\begin{figure}
\begin{center}
%\psbox[width=6.0cm]{}
\includegraphics[width=7.0cm]{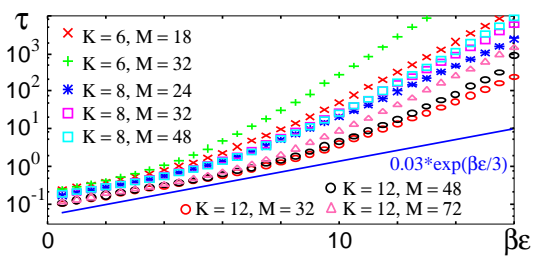}
\end{center}
\caption{Relaxation time as a function of $\beta$ for the sample reaction 
networks in Fig. 1(a)(b).}
\end{figure}

In Fig. 1, we give examples of the relaxation time course for four sets 
of networks ($M = 24$, $K = 8$), where we plot the deviation from equilibrium
concentration defined by
$C(t) = \{\sum_i (x_i(t) - x_i^{eq})(x_i(0) - x_i^{eq})\}/\{\sum_i(x_i(0) - x_i^{eq})^2\}$.
We note two salient behaviors when $\beta$ is sufficiently
larger than $\beta_c \sim 3/\varepsilon$, which is the inverse of the average difference between energy levels. First, there exists overall $\log(t)$ relaxation, in contrast to
exponential relaxation for small $\beta$.
Second, there are several plateaus in the relaxation course.  
The logarithmic relaxation at large $\beta$ is generally observed,
independently of the networks or K.  
Existence of plateaus is also universal. The number of plateaus 
depends on each network (Fig. 1(b)); generally, the number decreases as 
$K$ increases. 

The integrated relaxation time  \begin{math} \tau = <\int_0^{\infty} |C(t)| dt> \end{math} is 
plotted as a function of $\beta$ for several $M$ and $K$ in Fig. 2,
where $<\cdots >$ is the average over networks with given $M$ and $K$.
For the small $\beta$ regime in which $C(t)$ decays exponentially, $\tau$  
follows $\exp(\beta/\beta_c)$. It is 
the inverse of the average reaction rate to increase the energy, which 
gives the order of the relaxation time\cite{note_r}. 
For large $\beta$ giving 
$\log(t)$ relaxation, $\tau$ follows $\exp( R \beta \varepsilon)$ with $R$ approaching 
a larger constant with the increase in 
$\beta$. $R$ increases with the increase (decrease) in $M$ ($K$), respectively.

The $\log(t)$ relaxation with plateaus is often observed in glass theory 
and experiments\cite{GR1,KCM1}. In the present case, these relaxation
characteristics are partially explained by a rough estimate of the 
eigenvalue distribution in linear stability analysis.  Consider deviation 
from the equilibrium concentration as
$x_i(t)=x^{eq}_i+\delta x_i(t)$, where the equilibrium concentration
$x_i^{eq}= s \exp(-\beta E_i)$ is the fixed 
point solution of eq. (2).  By linearizing with 
$\delta x_i(t)$ ($i=1,\cdots,M$), we get ${\bf \delta \dot{x}(t)} = \bf{J}{\bf \delta x(t)}$ with the Jacobi matrix $\bf{J}$ computed straightforwardly.
For large $\beta$, $J_{i,j}$ for $i > j$,  given by $Con(i, j; c') x_{c'}^{eq}e^{-\beta(E_i - E_j)}$, is much smaller than that for 
$i < j$, $Con(i, j; c") x_{c"}^{eq}$.  If the former terms are neglected,  
the above $\bf{J}$ is a triangular matrix, so that the eigenvalues 
$\lambda_i$ of $\bf{J}$ are given by diagonal elements $J_{i,i}= -\sum_{j<i} Con(i, j; c") x_{c"}^{eq} -\sum_{j>i} Con(i, j; c') x_{c'}^{eq}e^{\beta(E_i - E_j)} $, 
whose distribution has  similar dependence to that of $\exp(-\beta E_k)$, 
for large $\beta$.  This is also true for the neglected off-diagonal terms. 
Hence, it is expected that the distribution of the eigenvalues $\lambda _i$ 
is similar to the distribution of $\exp(-\beta E_k)$, for large $\beta$ 
(besides the null eigenvalue $\lambda_0=0$ corresponding to the equilibrium 
distribution).  In fact, numerical diagonalization of the Jacobian matrix 
supports this estimate of eigenvalue distribution. By using this linear 
approximation and the correspondence of the eigenvalue with 
$\exp(-\beta E)$, $C(t)$ is approximated by 
$\int _0^{\varepsilon} D(E) a(E) \exp(-e^{-\beta E} t) dE$, with
the distribution of energy $D(E)$, which is roughly homogeneous, and 
the fractions of the eigenmodes $a(E)$ in the initial condition, which are 
almost equal. Hence, $D(E)$ and $a(E)$ are roughly constant\cite{note3}. 
By setting $u = \exp(-\beta E) t$, the integral is rewritten as
$(1/\beta)\int _{te^{-\beta \varepsilon}}^{t} (1/u)e^{-u} du$. 
By taking a limit of $\beta \to \infty$ first, $\log(t)$ dependence is 
obtained asymptotically for large $t$.

Though this estimate is originally asymptotic for large $t$, we used it
for the time span where many eigenvalues contribute to the relaxation.  For the last
stage of the relaxation, only a few eigenvalues contribute.  If there
is a gap $\Delta \lambda$ between two neighboring eigenvalues, there is a
plateau in the relaxation for the time span $\frac{1}{\Delta
\lambda}$.  For large $\beta$, the gap between eigenvalues increases
so that the existence of a plateau is expected.
However, plateaus other than the last one, as well as their number during 
the relaxation, are not directly obtained from this argument.  Here we give
a heuristic argument for the plateaus.  

\begin{figure}
\begin{center}
%\psbox[width=8.0cm]{}
\includegraphics[width=8.0cm]{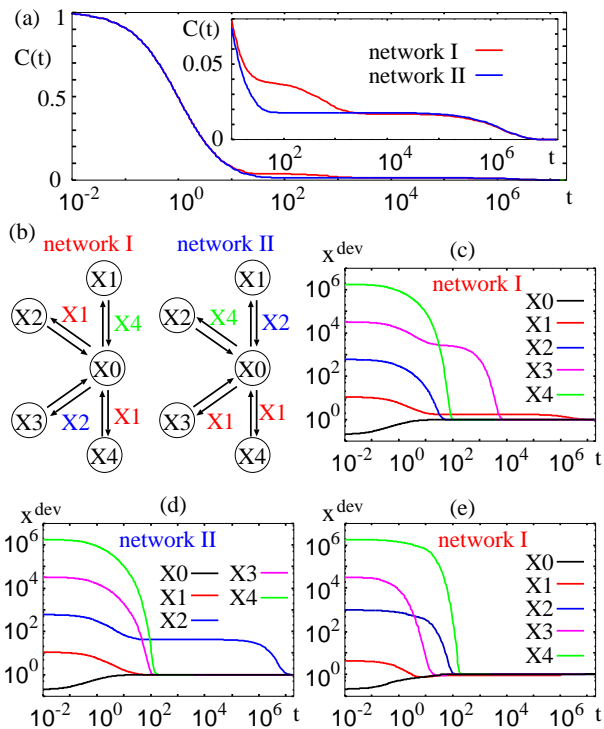}
\end{center}
\caption{(a) Time course of $C(t)$ for two example networks,  
I and II, given in (b), with $\beta = 16 /\varepsilon$. 
In (b), the chemicals attached to the arrows that display reactions are their catalysts.
Time courses of $x^{dev}_i(t)$ of (c) network I, 
and (d) network II corresponding to (a). In (a)(c)(d), all the chemicals are
at equal number initially (i.e., $\beta=0$). Time courses
of $x^{dev}_i(t)$ of network I from the initial condition
$x_i =1$ for $i = 0, 2, 3$, $x_1 = 0.4$ and $x_4 = 1.6$ are plotted
in (e).
}
\end{figure}

In Figs. 3 and 4, we give examples of the relaxation for smaller networks.
Besides $C(t)$, we have plotted $x^{dev}_i(t)=x_i(t)/x^{eq}_i$ in Figs. 3(c)-(e) 
and 4(c). At each plateau, there are cluster(s) of elements in which 
$x^{dev}_i(t)$ takes almost the same value.  Within each cluster, chemicals are 
in local equilibrium through mutual reactions, whereas the equilibration 
process with elements out of the cluster is suppressed, since the 
concentrations of the catalytic components responsible for reactions for such 
equilibration are low. 
Consider a chemical with $x^{dev}_i$ larger than the others. If the 
concentration of the catalyst(s) necessary to equilibrate the abundant 
chemical is small, the equilibration process is suppressed.  Negative 
correlation in the abundances between the excess chemical and its catalyst
will further suppress the relaxation to equilibrium. We now 
illustrate how this negative correlation gives rise to plateaus consisting 
of local-equilibrium clusters, by using examples given in Figs. 3 and 4.

In the networks I and II in Fig. 3(b), consisting of  5 chemicals, the component 
$X_0$ (with lowest $E$) is transformed to all other components. In cases with 
large $\beta$, because $E_0$ is minimum, chemicals $i \ge 1$ flow into $X_0$ 
from the initial condition 
with $\beta=0$, having $x^{dev}_i(0) > 1$ for $i \ge 1$ for large $\beta$.
For both the networks, the eigenvalues of $\bf{J}$ are $\exp(-\beta E_1)$, 
$\exp(-\beta E_2)$, $\exp(-\beta E_4)$, and 0 ($E_i = \frac{i}{4} \varepsilon$),
asymptotically as $\beta$ becomes large. As shown in Fig. 3(a), however, the 
numbers of plateaus appearing through the relaxation are different between the two 
networks.

In network I, the first plateau consists of a local-equilibrium cluster 
$X_0$, $X_2$, and $X_4$, whereas $X_3$ joins to the cluster at the second 
plateau, as shown in  Fig. 3(c).  The suppression of equilibration of $X_1$ 
is explained as follows: Relaxation (i.e., decrease) 
of $X_1$ ($X_4$) is catalyzed by $X_4$ ($X_1$), respectively.  If 
one of the species $X_1$ or $X_4$ decreases faster, the relaxation of 
the other is suppressed. Because $x^{eq}_1$ is larger than $x^{eq}_4$, $X_4$ 
relaxes faster, so that the relaxation of $X_1$ is suppressed.  The 
negative correlation between the abundances of $X_1$ and its catalyst 
hinders the relaxation of $X_1$. Since the 
relaxations of $X_2$ and $X_4$ are catalyzed by the abundant $X_1$, the
local-equilibrium $X_0$, $X_2$, and $X_4$ is first achieved and then $X_3$
catalyzed by $X_2$ (more abundant than $X_4$, the catalyst for $X_1$) joins 
to the cluster.

In network II, on the other hand, the relaxation of $X_1$ is not 
suppressed since its catalyst $X_2$ relaxes only slowly because its catalyst 
$X_4$ relaxes faster, as it is catalyzed by abundant $X_1$.  Negative 
correlation exists, not between $X_1$ and its catalyst, but instead 
between $X_2$ and its catalyst $X_4$.  Thus, the local equilibrium 
among $X_0$, $X_1$, $X_3$, and $X_4$ is realized to produce only one plateau.

As expected from the above argument, the types of plateaus that appear in the 
relaxation can depend on the initial condition, because the reactions 
that are suppressed depend on which catalysts are first decreased. See 
Fig. 3(e), which shows the relaxation process of network I from the 
initial condition with $x_4=4x_1$.

\begin{figure}
\begin{center}
%\psbox[width=8.0cm]{}
\includegraphics[width=8.0cm]{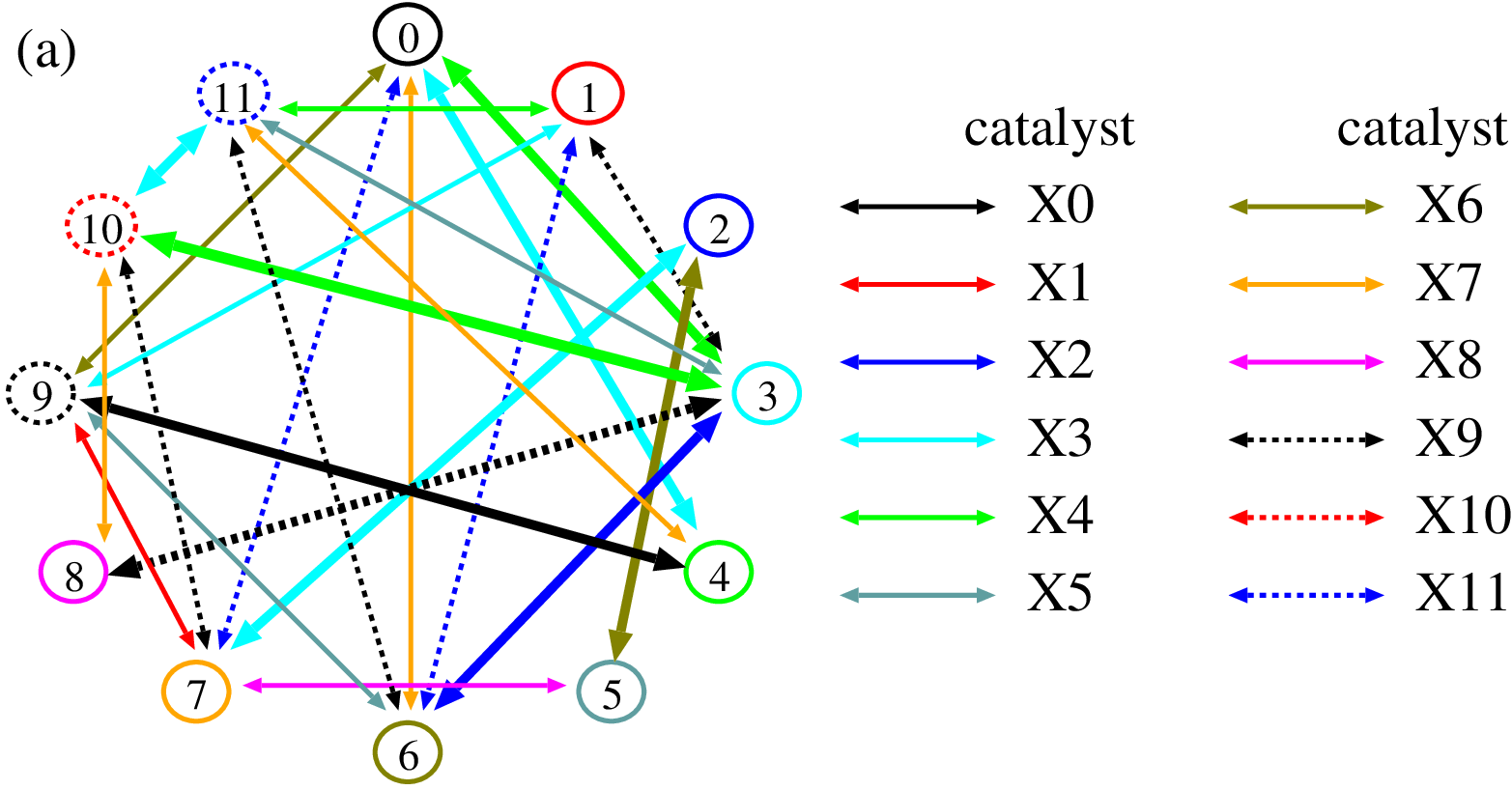}
\includegraphics[width=8.0cm]{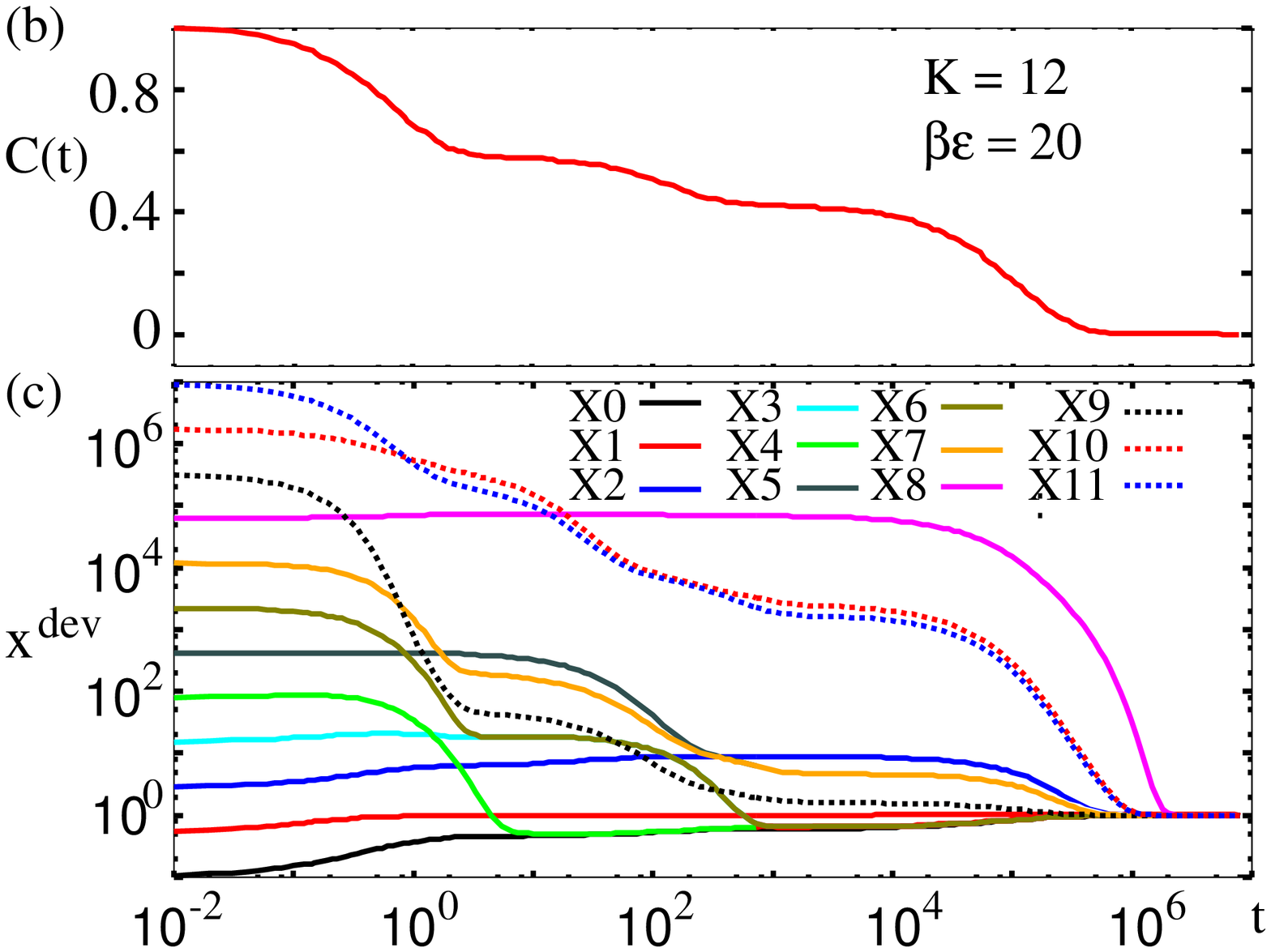}
\end{center}
\caption{(a) A reaction network with $M=12$ and $K=4$, where the color of 
each arrow shows the catalyst for the reaction, and thick arrows indicate 
''major relaxation'' for each chemical (see text).
(b) Time course of $C(t)$ for the network, and (c) time courses of 
$x^{dev}_i(t)$ for $\beta = 20 /\varepsilon$.}
\end{figure}
 
For complex catalytic reaction networks, the argument is not so simple, but
the existence of local equilibria and suppression of relaxation by
the negative correlation mechanism generally underlie the formation of 
plateaus. Figure 4(a) is a catalytic reaction network with $M = 12$ and 
$K=4$, and Fig. 4(b)(c) show the time courses of 
$C(t)$ and $x^{dev}_i(t)$ for $\beta = 20 /\varepsilon$.  As shown in 
Fig. 4(b), this network exhibits three plateaus in the relaxation process. 
At each plateau, chemicals $i=1,\cdots M$ are clustered into a few groups
within which $x^{dev}_i$ is almost constant (Fig. 4(c)). The following 
clusters are formed successively: $\{X_0,X_4\}$, $\{X_3,X_6\}$, and 
$\{X_{10},X_{11}\}$ at the first plateau, $\{X_0,X_3,X_4,X_6\}$, $\{X_5,X_7\}$, 
and $\{X_{10},X_{11}\}$ at the second plateau, and $\{X_0,X_1,\cdots,X_7,X_9\}$ 
and $\{X_{10},X_{11}\}$ at the third plateau. 

Each of these plateaus is explained by checking if the catalyst for the 
``majorh relaxation process for each $X_i$ is abundant, which is a path 
catalyzed by $X_k$ with the smallest $k$ among the reactions with $j$ 
smaller than $i$. At the 
first plateau, $X_3$ and $X_4$ have negative correlation since the major 
relaxation of $X_3$ is the reaction catalyzed by $X_4$, and that of $X_4$ 
by $X_3$, so that the formation of the cluster $\{X_0, X_4\}$ 
suppresses the equilibration between $X_0$ and $X_3$.  

At the second plateau, $\{X_3,X_6\}$ and $\{X_5,X_7\}$ clusters have 
the following negative correlation. For $\{X_5,X_7\}$ clusters, the 
reactions $X_5 + X_6 \to X_2 + X_6$ and $X_7 + X_3 \to X_2 + X_3$ give 
the major relaxations. On the other hand, the reactions 
$X_6 + X_7 \to X_0 + X_7$ and $X_3 + X_4 \to X_3 + X_4$ give the major 
relaxations for the $\{X_3,X_6\}$ cluster, but the reaction 
$X_3 + X_4 \to X_3 + X_4$ is suppressed since its catalyst $X_4$ has 
already been decreased. In this case, the cluster$\{X_5,X_7\}$ 
does not join $X_2$, whereas the clusters $\{X_3,X_6\}$ and $\{X_0,X_4\}$ 
aggregate.

In general, among a variety of chemical components, there exists
such a negative correlation between chemicals in excess and the catalysts
to decrease them towards equilibrium. Then, the
equilibration of the chemicals is suppressed, leading to a
plateau in the relaxation process.

In this Letter, slow relaxation to equilibrium in catalytic reaction 
networks is demonstrated. When the temperature of the system is sufficiently
lower than $\varepsilon$/3, the average difference between energy levels, 
overall $\log(t)$ relaxation appears. Several plateaus appear 
depending on the network and initial condition. 
The plateaus are not metastable states in the energy landscape but, rather, 
are a result of kinetic constraints due to a reaction bottleneck,
originating in the formation of local-equilibrium clusters and suppression of
equilibration by the negative correlation between an excess chemical and
its catalyst.

Possible configurations for local-equilibrium clusters are limited, and 
thus the number and ordering of plateaus are restricted. However, they 
are not necessarily uniquely determined by the network, but depend on 
the initial condition, because they are influenced by which catalysts are 
decreased first.  Also, the relaxation is often non-monotonic; the deviation 
from equilibrium may increase during the relaxation course. 
Such roundabout relaxation has also been observed in a Hamiltonian system
\cite{Morita-round}.
We also note that discreteness in the molecule number results in anomalous 
reaction dynamics with long time correlations\cite{AK2}, and further suppresses 
the relaxation in the catalytic reaction network\cite{inprep}.

The behaviors reported here are reminiscent of the relaxation in glass.  
Our model, as studied here, has a kinetic constraint, although the 
constraint is based on the network structure rather than the spatial configuration.  
Application of theoretical frameworks developed in the study of glasses will 
be important to our chemical net glass in future work.
Maintenance of the quasi-stationary states reported here, as well as successive changes in
them, are often observed in biochemical processes,
which have a large variance of reaction rates, i.e., potentiality of 
$\varepsilon \gg 1/\beta$. In future work, it will be important to discuss the relevance of 
the present ``glassyh dynamics to intracellular reactions.

The authors would like to thank M. Tachikawa, S. Sano, and S. Ishihara
for discussions. (A. A.) This research was supported in part by a Grant-in 
Aid for Young Scientist (B) (Grant No. 19740260).


\begin{thebibliography}{999}
\bibitem{note1}
We can adopt other forms of $k_{i,j}$ satisfying the detailed balance,
say $k_{i,j}=\exp(-\beta E_j)$.  Overall qualitative
behaviors -- $\log(t)$ relaxation and existence of plateaus -- are not altered.
\bibitem{note2}
Qualitatively identical behaviors are obtained even for a Gaussian distribution of energy levels, or  
bounded distributions.  For log-normal or power-law distributions of energy, the relaxation behavior
is altered.
\bibitem{note_r}
By linearizing the relaxation dynamics to equilibrium (as discussed later),
and by replacing the energy difference between chemicals by the
average $\varepsilon/3$,
we get $\tau \propto \exp(\beta \varepsilon /3)$ under suitable approximation.
\bibitem{note3} As long as there is no singular dependence of $D(E)$ and $a(E)$ on $E$ (such as the power law dependence) the estimate below is valid.
\end{thebibliography}
\end{document}